\documentclass[11pt, a4paper, fleqn]{article}

\usepackage[tbtags]{amsmath}
\usepackage{amssymb}
\usepackage{mathrsfs}

\newcommand{\ui}{\mathrm{i}\;\!}
\newcommand{\ud}{\mathrm{d}}
\newcommand{\uD}{\mathrm{D}}
\newcommand{\uU}{\mathrm{U}}
\newcommand{\uO}{\mathrm{O}}
\newcommand{\nc}[1]{\widehat{#1}}

\begin{document}


\title{\textbf{Seiberg--Witten maps and commutator anomalies\\ in noncommutative electrodynamics}}

\author{\textbf{Rabin Banerjee and Kuldeep Kumar}\\[0.5ex]
  \textit{\normalsize S.~N.~Bose National Centre for Basic Sciences,}\\[-0.2ex]
  \textit{\normalsize JD Block, Sector 3, Salt Lake, Kolkata 700098, India}\\
  {\normalsize E-mail: \texttt{rabin@bose.res.in}, \texttt{kuldeep@bose.res.in}}}

\date{}

\maketitle

\begin{abstract}
We exploit the Seiberg--Witten maps for fields and currents in a $\uU(1)$ gauge theory relating the noncommutative and commutative (usual) descriptions to obtain the $\uO(\theta)$ structure of the commutator anomalies in noncommutative electrodynamics. These commutators involve the (covariant) current--current algebra and the (covariant) current--field algebra. We also establish the compatibility of the anomalous commutators with the noncommutative covariant anomaly through the use of certain consistency conditions derived here.
\begin{center}{PACS: {11.10.Nx, 11.15.-q}; \quad  Appearing in \textit{Phys.~Rev.}~D}\end{center}

\end{abstract}

\bigskip


\section{\label{sec:intro} Introduction}

The subject of anomalies in gauge theories has been studied extensively in the literature~\cite{Bertlmann:2001}. Ever since the importance of noncommutative manifolds was realised~\cite{Douglas:2001}, it has been natural to investigate the structure of anomalies in such a setting. Various results have been reported in this context. In particular, it has been noted~\cite{Ardalan:2000} that, due to noncommutativity, two different currents can be defined even for a $\uU(1)$ theory. These are the star-gauge-invariant and the star-gauge-covariant currents which are defined according to their distinct gauge-transformation properties. In this paper we shall be exclusively dealing with the star-gauge-covariant currents. Now the covariant divergence of the star-gauge-covariant axial current reveals an anomaly---this is the star-gauge-covariant anomaly~\cite{AS-GM} which is basically the covariant deformation of the usual gauge-invariant Adler--Bell--Jackiw (ABJ) anomaly~\cite{Adl}.

The next logical step would be to compute the anomalous commutators involving the currents and see their connection with the anomaly, as happens for the commutative description~\cite{AB, Jackiw:1968}. The structure of the anomalous commutators in the noncommutative setting, however, is lacking in the literature. It is the aim of this paper to investigate this aspect. Here we would like to mention that the computation of noncommutative commutators from loop diagrams following the `Bjorken-limit' approach might not be practically feasible. Even in the ordinary case, the computation of anomalous commutators is much more involved than that of the divergence anomaly.

Here we provide an approach to obtain the structure of the anomalous commutators in a noncommutative theory. We exploit the maps for fields and currents in a $\uU(1)$ gauge theory in noncommutative and commutative (usual) descriptions~\cite{SW, BLY, BK} to express the commutators in the noncommutative theory in favour of their commutative counterparts, where the results are known~\cite{AB}. Using these known results we obtain the explicit structure for the anomalous commutators in the noncommutative theory.

The new results on anomalous commutators in noncommutative electrodynamics are by themselves interesting. Their compatibility with the noncommutative divergence anomalies, exhibited through consistency conditions derived here, further supports our analysis. Moreover the computational method provides a nontrivial application of various Seiberg--Witten (SW) maps.

The paper is organised as follows. In Sec.~\ref{sec:review} we briefly review the maps for currents and anomalies in a $\uU(1)$ gauge theory in the two descriptions. These maps are later used to express the anomalous commutators in noncommutative electrodynamics in terms of their commutative counterparts. After enumerating the known results for ordinary anomalous commutators in the first part of Sec.~\ref{sec:anom-com}, we compute the commutators in the noncommutative theory in the second part. Although we have considered massless quantum electrodynamics (QED) here, the structure of these commutators remains equally valid for the massive case as well. Explicit results are given for the current--current as well as the current--field commutators. The compatibility of our results for these anomalous commutators with the noncommutative covariant anomaly has been established in Sec.~\ref{sec:cons} through the use of certain consistency conditions. Finally, it is known that in the ordinary theory there is a possibility of the presence of additional terms in some of the commutators. Section \ref{sec:ambig} deals with the implications of these ambiguities on our scheme. Our conclusions are left for Sec.~\ref{sec:conclu}.


\section{\label{sec:review}A brief review of the maps for currents and anomalies}

In order to set up notations and make the paper self-contained, we briefly review here the maps relating the currents and anomalies in the noncommutative and commutative (usual) descriptions. We shall restrict to the first order in $\theta$, the noncommutativity parameter. The original maps \cite{SW} involving the gauge potentials, field tensors and gauge parameters in a $\uU(1)$ gauge theory in the two descriptions are given by\footnote{Here the hat-variables refer to the variables in noncommutative space.}
\begin{gather}
\label{A.m}
\nc{A}_{\mu} = A_{\mu}-\frac{1}{2}\theta^{\alpha\beta}A_{\alpha}\left(\partial_{\beta}A_{\mu}+F_{\beta\mu}\right)+\uO(\theta^{2}),\displaybreak[0]\\
\label{F.m}
\nc{F}_{\mu\nu} = F_{\mu\nu}-\theta^{\alpha\beta}\left(A_{\alpha}\partial_{\beta}F_{\mu\nu}+F_{\mu\alpha}F_{\beta\nu}\right)+\uO(\theta^{2}),\displaybreak[0]\\
\label{lamb.m}
\nc{\lambda} = \lambda-\frac{1}{2}\theta^{\alpha\beta}A_{\alpha}\partial_{\beta}\lambda+\uO(\theta^{2}),
\end{gather}
which ensure the stability of gauge transformations
\begin{gather}
\label{A-hat.gt}
\nc{\delta}_{\nc{\lambda}}\nc{A}_{\mu} = \nc{\uD}_{\mu}\star\nc{\lambda}
\equiv\partial_{\mu}\nc{\lambda}+\ui\left[\nc{\lambda},\nc{A}_{\mu}\right]_{\star}
= \partial_{\mu}\nc{\lambda}+\theta^{\alpha\beta}\partial_{\alpha}\nc{A}_{\mu}\partial_{\beta}\nc{\lambda}+\uO(\theta^{2}),\\
\label{A.gt}
\delta_{\lambda}A_{\mu} = \partial_{\mu}\lambda,
\end{gather}
where $[\nc{\lambda},\nc{A}_{\mu}]_{\star} \equiv \nc{\lambda}\star\nc{A}_{\mu}-\nc{A}_{\mu}\star\nc{\lambda}$, and the star product of two fields $A(x)$ and $B(x)$ is defined as
\begin{equation}\label{star.d}
(A\star B)(x)=\left.\exp\left(\frac{\ui}{2}\theta^{\alpha\beta}\partial_{\alpha}\partial'_{\beta}\right)A(x)B(x')\right|_{x'=x}.
\end{equation}
The map (\ref{F.m}) is a consequence of the map (\ref{A.m}) following from the basic definitions
\begin{gather}
\label{F-hat.d}
\nc{F}_{\mu\nu}=\partial_{\mu}\nc{A}_{\nu}-\partial_{\nu}\nc{A}_{\mu}-\ui\left[\nc{A}_{\mu},\nc{A}_{\nu}\right]_{\star}
=\partial_{\mu}\nc{A}_{\nu}-\partial_{\nu}\nc{A}_{\mu}+\theta^{\alpha\beta}\partial_{\alpha}\nc{A}_{\mu}\partial_{\beta}\nc{A}_{\nu}+\uO(\theta^2),\\
\label{F.d}
F_{\mu\nu}=\partial_{\mu}A_{\nu}-\partial_{\nu}A_{\mu},
\end{gather}
so that, whereas $F_{\mu\nu}$ is gauge invariant, $\nc{F}_{\mu\nu}$ transforms covariantly under the star-gauge transformation:
\begin{equation}\label{F-hat.gt}
\nc{\delta}_{\nc{\lambda}}\nc{F}_{\mu\nu} = \ui\left[\nc{\lambda},\nc{F}_{\mu\nu}\right]_{\star} = \theta^{\alpha\beta}\partial_{\alpha}\nc{F}_{\mu\nu}\partial_{\beta}\nc{\lambda}+\uO(\theta^2).
\end{equation}

In order to discuss noncommutative gauge theories with sources, it is essential to have a map for the sources also, so that a complete transition between noncommutative gauge theories and the usual ones is possible. Let the noncommutative action be defined as
\begin{equation}\label{S-hat}
\nc{S}(\nc{A},\nc{\psi})
= -\frac{1}{4}\int\!\ud^{4}x\,\nc{F}_{\mu\nu}\star\nc{F}^{\mu\nu}+\nc{S}_{\mathrm{M}}(\nc{\psi},\nc{A})
\equiv \nc{S}_{\mathrm{ph}}(\nc{A})+\nc{S}_{\mathrm{M}}(\nc{\psi},\nc{A}),
\end{equation}
where the pure gauge term has been isolated in the `photonic' piece $\nc{S}_{\mathrm{ph}}(\nc{A})$, and $\nc{\psi}_{\alpha}$ are the charged matter fields. The equation of motion for $\nc{A}_{\mu}$ is
\begin{equation}\label{eom-nc}
\frac{\delta\nc{S}_{\mathrm{ph}}}{\delta\nc{A}_{\mu}}
= \nc{\uD}_{\nu}\star\nc{F}^{\nu\mu} = \nc{J}^{\mu},
\end{equation}
where
\begin{equation}\label{J-hat.d}
\nc{J}^{\mu}=-\left.\frac{\delta\nc{S}_{\mathrm{M}}}{\delta\nc{A}_{\mu}}\right|_{\nc{\psi}}.
\end{equation}
Equation (\ref{eom-nc}) shows that $\nc{J}^{\mu}$ is star-gauge covariant:
\begin{equation}\label{sgt2}
\nc{\delta}_{\nc{\lambda}}\nc{J}^{\mu}
= \ui\left[\nc{\lambda}, \nc{J}^{\mu}\right]_{\star}
= \theta^{\alpha\beta}\partial_{\alpha}\nc{J}^{\mu}\partial_{\beta}\nc{\lambda}+\uO(\theta^{2}),
\end{equation}
and satisfies the noncommutative covariant conservation law
$\nc{\uD}_{\mu}\star\nc{J}^{\mu}=0$.

The use of SW map in the action (\ref{S-hat}) gives its $\theta$-expanded version in commutative space:
\begin{equation}\label{S-theta}
\nc{S}(\nc{A},\nc{\psi}) \rightarrow S^{\theta}(A,\psi)
= S^{\theta}_{\mathrm{ph}}(A)+S^{\theta}_{\mathrm{M}}(\psi,A),
\end{equation}
where $S^{\theta}_{\mathrm{ph}}(A)$ contains all terms involving $A_{\mu}$ only, and is given by
\begin{equation}\label{S-th-ph}
S^{\theta}_{\mathrm{ph}} = -\frac{1}{4}\int\!\ud^{4}x\,\left[F_{\mu\nu}F^{\mu\nu}+\theta^{\alpha\beta}F^{\mu\nu}\left(2F_{\mu\alpha}F_{\nu\beta}+\frac{1}{2}F_{\beta\alpha}F_{\mu\nu}\right)+\uO(\theta^{2})\right].
\end{equation}
The equation of motion following from the action (\ref{S-theta}) is
\begin{equation}\label{eom-th}
\frac{\delta S^{\theta}_{\mathrm{ph}}}{\delta A_{\mu}}=J^{\mu},
\end{equation}
where
\begin{equation}\label{J.d}
J^{\mu}=-\left.\frac{\delta S^{\theta}_{\mathrm{M}}}{\delta A_{\mu}}\right|_{\psi}.
\end{equation}
Expectedly, from these relations it follows that $J^{\mu}$ is gauge invariant and satisfies the ordinary conservation law $\partial_{\mu}J^{\mu} = 0$.

The stability of the $\uO(\theta)$ map among the currents $\nc{J}^\mu$ and $J^\mu$ under gauge transformations is easily attained by mimicking the map \eqref{F.m} among the field tensors:
\begin{equation*}
\nc{J}^\mu = J^\mu - \theta^{\alpha\beta} A_\alpha\partial_{\beta}J^\mu + (\cdots)+\uO(\theta^2),
\end{equation*}
where $(\cdots)$ indicates the freedom of adding more $\uO(\theta)$ terms that are invariant under ordinary gauge transformations. It is clear that the most general structure is given by
\begin{equation*}
\nc{J}^\mu = J^\mu - \theta^{\alpha\beta} A_\alpha\partial_{\beta}J^\mu + c_{1}\theta^{\mu\alpha}F_{\alpha\beta}J^{\beta}+c_{2}\theta^{\alpha\beta}F_{\alpha\beta}J^\mu+ c_{3}\theta^{\alpha\beta}{F_\alpha}^{\mu}J_\beta+\uO(\theta^2),
\end{equation*}
where $c_1$, $c_2$ and $c_3$ are undetermined coefficients. Demanding the simultaneous conservation $\nc{\uD}_{\mu}\star \nc{J}^\mu = \partial_{\mu}J^\mu = 0$ immediately fixes $c_1 = 2c_2 = 1$ and $c_3 = 0$, so that
\begin{equation}\label{J-hat2.m}
\nc{J}^{\mu} = J^{\mu} - \theta^{\alpha\beta}\partial_{\beta}\left(A_{\alpha}J^{\mu}\right) + \theta^{\mu\alpha}F_{\alpha\beta}J^{\beta} + \uO(\theta^{2}).
\end{equation}
This is the $\uO(\theta)$ map among the currents obtained in an algebraic approach. The map can be generalised to higher orders in $\theta$ in a dynamical approach \cite{BLY}.

Using the maps (\ref{A.m}) and (\ref{J-hat2.m}), the covariant divergence of $\nc{J}^{\mu}$,
\begin{equation}\label{DJ}
\nc{\uD}_{\mu}\star\nc{J}^{\mu}
= \partial_{\mu}\nc{J}^{\mu}+\ui\left[\nc{J}^{\mu},\nc{A}_{\mu}\right]_{\star} = \partial_{\mu}\nc{J}^{\mu}-\theta^{\alpha\beta}\partial_{\alpha}\nc{J}^{\mu}\partial_{\beta}\nc{A}_{\mu}+\uO(\theta^{2}),
\end{equation}
can be expressed in terms of the ordinary divergence of $J^\mu$ as~\cite{BLY}
\begin{equation}\label{DJ2}
\nc{\uD}_{\mu}\star\nc{J}^{\mu}
= \partial_{\mu}J^{\mu}+\theta^{\alpha\beta}\partial_{\alpha}\left(A_{\beta}\partial_{\mu}J^{\mu}\right)+\uO(\theta^{2}),
\end{equation}
so that the covariant conservation of $\nc{J}^{\mu}$ follows from the ordinary conservation of $J^{\mu}$. (An analogous relation exists for non-Abelian groups also~\cite{BK}.)

The analysis presented above for the vector current can be readily taken over for the axial current. Classically everything would be fine since the relevant currents satisfy identical gauge-transformation properties and conservation laws as the vector current. At the quantum level, however, the axial currents are not conserved. The standard ABJ anomaly \cite{Adl} is not modified in $\theta$-expanded gauge theory \cite{BMR}, and is given by
\begin{equation}\label{anom.d}
\partial_{\mu}J^{\mu}_{5} = \mathscr{A} \equiv \frac{1}{16\pi^{2}}\varepsilon_{\mu\nu\lambda\rho}F^{\mu\nu}F^{\lambda\rho},
\end{equation}
whereas the star-gauge-covariant anomaly is just a standard deformation of the above result~\cite{AS-GM}:
\begin{equation}\label{anom-cov.d}
\nc{\uD}_{\mu}\star\nc{J}^{\mu}_{5} = \nc{\mathscr{A}} \equiv \frac{1}{16\pi^{2}}\varepsilon_{\mu\nu\lambda\rho}\nc{F}^{\mu\nu}\star\nc{F}^{\lambda\rho}.
\end{equation}
The expected map for anomalies, obtained by a lift from the classical result (\ref{DJ2}), follows as
\begin{equation}\label{anom.m}
\nc{\mathscr{A}} = \mathscr{A} + \theta^{\alpha\beta}\partial_{\alpha}\left(A_{\beta}\mathscr{A}\right) + \uO(\theta^{2}).
\end{equation}
It has been shown~\cite{BK} that the $\uO(\theta)$ map (\ref{anom.m}) is indeed valid, although derivative corrections are needed at higher orders. This implies that, up to $\uO(\theta)$, the source map (\ref{J-hat2.m}) can be used to relate the axial currents at the quantum level as well, so that
\begin{equation}\label{J-hat5.m}
\nc{J}^{\mu}_{5} = J^{\mu}_{5} - \theta^{\alpha\beta}\partial_{\beta}\left(A_{\alpha}J^{\mu}_{5}\right) + \theta^{\mu\alpha}F_{\alpha\beta}J^{\beta}_{5} + \uO(\theta^{2}).
\end{equation}
We shall subsequently exploit the maps \eqref{J-hat2.m} and \eqref{J-hat5.m} to determine the various anomalous commutators in noncommutative electrodynamics.


\section{\label{sec:anom-com}Anomalous commutators}

Our method of computing the commutators is straight-forward. The maps connecting the variables in the two descriptions will be used to express the commutators in the noncommutative theory in favour of their commutative counterparts. From a knowledge of the latter the former is easily obtained. Let us thus enumerate the various anomalous commutators in the ordinary theory.

\subsection{Anomalous commutators in the ordinary theory}

We consider
massless QED given by the Lagrangian density\footnote{We use the $(+,-,-,-)$ signature, and take $\varepsilon_{0123}=\varepsilon_{123}=1$, $E^{i} = F_{0i}$, $B^{i} = -\varepsilon_{ijk}\partial_{j}A_{k}$ with $i,j,k = 1, 2, 3$.}
\begin{equation}\label{S101}
\mathscr{L} = \ui\bar{\psi}\gamma^{\mu}\partial_{\mu}\psi-\frac{1}{4}F_{\mu\nu}F^{\mu\nu}-\bar{\psi}\gamma^{\mu}\psi A_{\mu}.
\end{equation}
The equations of motion for the fields are
\begin{gather}
\label{Seom1}
\ui\gamma^{\mu}\partial_{\mu}\psi = \gamma^{\mu}\psi A_{\mu},\\
\label{Seom2}
\partial_{\nu}F^{\nu\mu} = J^{\mu},
\end{gather}
where $J^{\mu} = \bar{\psi}\gamma^{\mu}\psi$. The usual current conservation, $\partial_{\mu}J^{\mu} = 0$, follows upon using the equation of motion. The canonical anticommutator relations of the spinor fields are
\begin{equation}\label{Sacr}
\left\{\psi_{\alpha}(\mathbf{x},t), \psi^{\dagger}_{\beta}(\mathbf{y},t)\right\}
= \delta_{\alpha\beta}\delta^{3}(\mathbf{x}-\mathbf{y}),
\end{equation}
with $\alpha, \beta = 1, \ldots, 4$, the labels of the spinor components, and the canonical commutation relations of the photon fields in the Feynman gauge are
\begin{equation}\label{Scr}
\begin{aligned}
&\left[A_{\mu}(\mathbf{x},t), \partial_{0}A_{\nu}(\mathbf{y},t)\right] = -\ui\eta_{\mu\nu}\delta^{3}(\mathbf{x}-\mathbf{y}),\\
&\left[A_{\mu}(\mathbf{x},t), A_{\nu}(\mathbf{y},t)\right] = \left[\partial_{0}A_{\mu}(\mathbf{x},t), \partial_{0}A_{\nu}(\mathbf{y},t)\right] = 0.
\end{aligned}
\end{equation}
It has been shown \cite{Adl} that the axial-vector current does not satisfy the usual divergence equation $\partial_{\mu}J^{\mu}_{5} = 0$ expected from naive use of equations of motion.\footnote{Whether the index `5' appears as a subscript or as a superscript is a matter of notational convenience: $J^{\mu}_{5}= \bar{\psi}\gamma^{\mu}\gamma_{5}\psi$, $J_{\mu}^{5}= \bar{\psi}\gamma_{\mu}\gamma_{5}\psi$.} Rather it satisfies the anomalous divergence equation given by Eq.~(\ref{anom.d}). The commutators\footnote{All the commutators appearing in this paper are equal-time commutators. By $[J_{0}(x), J_{0}^5(y)]$ we mean $[J_{0}(\mathbf{x},t), J_{0}^5(\mathbf{y},t)]$, and so on. Likewise, $S_{00}(x,y)$ appearing in Eq.~\eqref{CS101} is to be understood as $S_{00}(\mathbf{x}, \mathbf{y}, t)$, and similarly for others.} involving the axial current which are relevant in the present context are \cite{AB}
\begin{gather}
\label{CS101}
S_{00}(x,y) \equiv \left[J_{0}(x), J_{0}^5(y)\right] = \frac{\ui}{4\pi^2}\varepsilon_{ijk}F_{jk}(y)\partial_{i}^x\delta^{3}(\mathbf{x}-\mathbf{y}),\displaybreak[0]\\
\label{CS102}
S_{i0}(x,y) \equiv \left[J_{i}(x), J_{0}^5(y)\right] = -\frac{\ui}{4\pi^2}\varepsilon_{ijk}F_{0j}(x)\partial_{k}^y\delta^{3}(\mathbf{x}-\mathbf{y}),\displaybreak[0]\\
\label{CS103}
S_{0i}(x,y) \equiv \left[J_{0}(x), J_{i}^5(y)\right] = \frac{\ui}{4\pi^2}\varepsilon_{ijk}F_{0j}(y)\partial_{k}^x\delta^{3}(\mathbf{x}-\mathbf{y}),\displaybreak[0]\\
\label{CA101}
L_{\sigma\mu}(x,y) \equiv \left[A_{\sigma}(x), J_{\mu}^5(y)\right] = 0,\displaybreak[0]\\
\label{CB101}
M_{0\mu}(x,y) \equiv \left[\partial_{0}A_{0}(x), J_{\mu}^5(y)\right] = 0,\displaybreak[0]\\
\label{CB102}
M_{i0}(x,y) \equiv \left[\partial_{0}A_{i}(x), J_{0}^5(y)\right] = \frac{\ui}{4\pi^2}\varepsilon_{ijk}F_{jk}\delta^{3}(\mathbf{x}-\mathbf{y}),\displaybreak[0]\\
\label{CB103}
M_{im}(x,y) \equiv \left[\partial_{0}A_{i}(x), J_{m}^5(y)\right] = \frac{\ui}{4\pi^2}\varepsilon_{imn}F_{0n}\delta^{3}(\mathbf{x}-\mathbf{y}).\end{gather}
All of the nonvanishing commutators given above are anomalous in the sense that if they are calculated by naive use of canonical commutation relations they vanish. These brackets are compatible with the axial anomaly (\ref{anom.d}) as shown in Ref.~\cite{AB}. Some other commutators which will be useful later are
\begin{equation}\label{SCC}
\left[J_{\mu}(x), A_{\sigma}(y)\right]
= \left[J_{0}(x), \partial_{0}A_{\sigma}(y)\right] = 0.
\end{equation}

\subsection{Anomalous commutators in the noncommutative theory}

Now we are in a position to compute the anomalous commutators in the noncommutative theory. In the context of the ordinary theory it is well-known that the anomalous commutators are a different manifestation of the ABJ anomly. Since the standard ABJ anomaly is not modified in $\theta$-expanded theory, we argue that the set \eqref{CS101}--\eqref{CB103} of commutators remains valid in the $\theta$-expanded theory also. We further note that the equation of motion for the photon field in $\theta$-expanded theory will differ from \eqref{Seom2} by an $\uO(\theta)$ term. This will modify the canonical commutation relation $[A_{i}(x), \partial_{0}A_{j}(y)]$ given in Eq.~\eqref{Scr}, which will have an $\uO(\theta)$ extension. But we need not compute this $\uO(\theta)$ correction explicitly since later we shall use this particular commutation relation in such terms which will already involve $\theta$. The commutators $[A_{0}(x), \partial_{0}A_{\nu}(y)]$ and $[A_{\mu}(x), \partial_{0}A_{0}(y)]$ will not have any $\uO(\theta)$ extension.

Although our main interest is in the current--current commutators, we shall compute some other commutators as well which will later be useful when we discuss the consistency conditions. Now onwards we shall take $\theta$ to be of `magnetic' type so that $\theta^{0i} = 0$. Using the maps \eqref{J-hat2.m}, \eqref{J-hat5.m}, and Eq.~\eqref{SCC}, we find
\begin{equation}\label{CCA1011}\begin{split}
\nc{S}_{00}(x,y) &\equiv \left[\nc{J}_{0}(x), \nc{J}_{0}^5(y)\right]\\
&= S_{00}(x,y) - \theta^{mn}\left[\partial_{n}^{y}\left(A_{m}(y)S_{00}(x,y)\right)\right.\\
&\qquad\qquad\qquad\qquad \left.{} + \partial_{n}^{x}\left(A_{m}(x)S_{00}(x,y) + J_{0}(x)L_{m0}(x,y)\right)\right] + \uO(\theta^2),
\end{split}\end{equation}
which may also be interpreted as a SW-type map. Proceeding similarly, we obtain\footnote{To save space we omit arguments, writing $S_{\mu\sigma}$, $L_{\mu\sigma}$ and $M_{\mu\sigma}$ instead of $S_{\mu\sigma}(x,y)$, $L_{\mu\sigma}(x,y)$ and $M_{\mu\sigma}(x,y)$ respectively.}
\begin{gather}
\label{CCA102}\begin{split}
\nc{S}_{i0}(x,y) &\equiv \left[\nc{J}_{i}(x), \nc{J}_{0}^5(y)\right]\\
&= S_{i0} - \theta^{mn}\left[\partial_{n}^{y}\left(A_{m}(y)S_{i0}\right) + \partial_{n}^{x}\left(A_{m}(x)S_{i0} + J_{i}(x)L_{m0}\right)\right]\\
&\quad {} - \theta^{im}\left[{F_{m}}^{\beta}(x)S_{\beta 0} + J_{0}(x)\left(\partial_{m}^{x}L_{00} - M_{m0}\right) - J_{n}(x)\left(\partial_{m}^{x}L_{n0} - \partial_{n}^{x}L_{m0}\right)\right]\\
&\quad {} + \uO(\theta^2),
\end{split}\\
\label{CSA103}\begin{split}
\nc{S}_{0i}(x,y) &\equiv \left[\nc{J}_{0}(x), \nc{J}_{i}^5(y)\right]\\
&= S_{0i} - \theta^{mn}\left[\partial_{n}^{y}\left(A_{m}(y)S_{0i}\right) + \partial_{n}^{x}\left(A_{m}(x)S_{0i} + J_{0}(x)L_{mi}\right)\right]\\
&\quad {} - \theta^{im}{F_{m}}^{\beta}(y)S_{0\beta} + \uO(\theta^2).
\end{split}
\end{gather}
The field--current algebra is likewise computed using Eqs.~\eqref{A.m}, \eqref{J-hat5.m} and \eqref{Scr}:
\begin{gather}
\label{CCA014}\begin{split}
\nc{L}_{00}(x,y) &\equiv \left[\nc{A}_{0}(x), \nc{J}_{0}^5(y)\right]\\
&= L_{00} - \theta^{mn}\left[\partial_{n}^{y}\left(A_{m}(y)L_{00}\right) + \frac{1}{2}A_{m}(x)\left(2\partial_{n}^{x}L_{00} - M_{n0}\right)\right.\\
&\qquad\qquad\qquad\quad \left.{} + \frac{1}{2}L_{m0}\left(\partial_{n}A_{0}(x)+F_{n0}(x)\right)\right]+\uO(\theta^2),
\end{split}\displaybreak[0]\\
\label{CCA014bb}
\begin{split}
\nc{L}_{0i}(x,y) &\equiv \left[\nc{A}_{0}(x), \nc{J}_{i}^5(y)\right]\\
&= L_{0i} - \theta^{im}{F_{m}}^{\beta}(y)L_{0\beta} - \theta^{mn}\left[\partial_{n}^{y}\left(A_{m}(y)L_{0i}\right) + \frac{1}{2}A_{m}(x)\left(2\partial_{n}^{x}L_{0i} - M_{ni}\right)\right.\\
&\qquad\qquad\qquad\qquad\qquad\qquad\qquad \left.{} + \frac{1}{2}L_{mi}\left(\partial_{n}A_{0}(x)+F_{n0}(x)\right)\right] + \uO(\theta^2),
\end{split}\displaybreak[0]\\
\label{CCA105}\begin{split}
\nc{L}_{i0}(x,y) &\equiv \left[\nc{A}_{i}(x), \nc{J}_{0}^5(y)\right]\\
&= L_{i0} - \theta^{mn}\bigg[\partial_{n}^{y}\left(A_{m}(y)L_{i0}\right) + \frac{1}{2}A_{m}(x)\left(2\partial_{n}^{x}L_{i0} - \partial_{i}^{x}L_{n0}\right)\\
&\qquad\qquad\qquad {} + \frac{1}{2}L_{m0}\left(\partial_{n}A_{i}(x)+F_{ni}(x)\right)\bigg] + \uO(\theta^2),
\end{split}\displaybreak[0]\\
\label{CCA1052} 
\begin{split}
\nc{L}_{im}(x,y) &\equiv \left[\nc{A}_{i}(x), \nc{J}_{m}^5(y)\right]\\
&= L_{im} + \ui\theta^{mi}J_{0}^{5}\delta^{3}(\mathbf{x}-\mathbf{y}) - \theta^{mj}{F_{j}}^{\beta}(y)L_{i\beta}\\
&\quad {} - \frac{1}{2}\theta^{jk}\left[2\partial_{k}^{y}\left(A_{j}(y)L_{im}\right) + L_{jm}\left(\partial_{k}A_{i}(x)+F_{ki}(x)\right)\right.\\
&\qquad\qquad\quad \left.{} + A_{j}(x)\left(2\partial_{k}^{x}L_{im}-\partial_{i}^{x}L_{km}\right)\right] + \uO(\theta^2),
\end{split}\displaybreak[0]\\
\label{CCA106} \raisetag{21pt}
\begin{split}
\nc{M}_{00}(x,y) &\equiv \left[\partial_{0}\nc{A}_{0}(x), \nc{J}_{0}^5(y)\right]\\
&= M_{00} - \theta^{mn}\bigg\{\partial_{n}^{y}\left(A_{m}(y)M_{00}\right) + \frac{1}{2}A_{m}(x)\left(2\partial_{n}^{x}M_{00} - \left[\partial_{0}\partial_{0}A_{n}(x),J_{0}^{5}(y)\right]\right)\\
&\qquad\qquad\qquad\quad {} + \frac{1}{2}L_{m0}\partial_{0}\left(\partial_{n}A_{0}(x)+F_{n0}(x)\right) + \partial_{0}A_{m}(x)\partial_{n}^{x}L_{00}\\
&\qquad\qquad\qquad\quad {} - \partial_{m}A_{0}(x)M_{n0}\bigg\}  + \uO(\theta^2),
\end{split}\displaybreak[0]\\
\label{CCA107}\begin{split}
\nc{M}_{i0}(x,y) &\equiv \left[\partial_{0}\nc{A}_{i}(x), \nc{J}_{0}^5(y)\right]\\
&= M_{i0} + \ui\theta^{in}\partial_{n}^y\left(J_{0}^5\delta^{3}(\mathbf{x}-\mathbf{y})\right)\\
&\quad {} - \theta^{mn}\bigg[\partial_{n}^{y}\left(A_{m}(y)M_{i0}\right) + \frac{1}{2}\partial_{0}A_{m}(x)\left(2\partial_{n}^{x}L_{i0}-\partial_{i}^{x}L_{n0}\right)\\
&\qquad\qquad\quad {}  + \frac{1}{2}L_{m0}\partial_{0}\left(\partial_{n}A_{i}(x)+F_{ni}(x)\right) + \frac{1}{2}\partial_{i}^{x}\left(A_{n}(x)M_{m0}\right)\\
&\qquad\qquad\quad {} + F_{ni}(x)M_{m0} + A_{m}(x)\partial_{n}^{x}M_{i0} \bigg] + \uO(\theta^2),
\end{split}\displaybreak[0]\\
\label{CCA108}\begin{split}
\nc{M}_{ik}(x,y) &\equiv \left[\partial_{0}\nc{A}_{i}(x), \nc{J}_{k}^5(y)\right]\\
&= M_{ik} + \ui\theta^{in}\partial_{n}^y\left(J_{k}^5\delta^{3}(\mathbf{x}-\mathbf{y})\right) + \ui\theta^{ki}J_{n}^{5}(y)\partial_{n}^{y}\delta^{3}(\mathbf{x}-\mathbf{y})\\
&\quad {} - \theta^{km}\left(\ui J_{i}^{5}(y)\partial_{m}^{y}\delta^{3}(\mathbf{x}-\mathbf{y}) + {F_{m}}^{\beta}(y)M_{i\beta}\right)\\
&\quad {} - \theta^{mn}\bigg[\partial_{n}^{y}(A_{m}(y)M_{ik}) + \frac{1}{2}\partial_{0}A_{m}(x)\left(2\partial_{n}^{x}L_{ik}-\partial_{i}^{x}L_{nk}\right)\\
&\qquad\qquad\quad {}  + \frac{1}{2}L_{mk}\partial_{0}\left(\partial_{n}A_{i}(x)+F_{ni}(x)\right) + \frac{1}{2}\partial_{i}^{x}\left(A_{n}(x)M_{mk}\right)\\
&\qquad\qquad\quad {} + F_{ni}(x)M_{mk} + A_{m}(x)\partial_{n}^{x}M_{ik}\bigg] + \uO(\theta^2).
\end{split}
\end{gather}
Now we use the relations \eqref{CS101}--\eqref{CB103} to substitute for the commutators appearing on the right-hand sides in Eqs.~\eqref{CCA1011}--\eqref{CCA108}. In order to compute $[\partial_{0}\partial_{0}A_{n}(x), J_{0}^5(y)]$ appearing on the right-hand side of Eq.~\eqref{CCA106}, we make use of the equation of motion. The equation of motion \eqref{Seom2} of the usual theory in the Feynman gauge reads $\partial_{0}\partial_{0}A_{\mu}-\boldsymbol{\nabla}^{2}A_{\mu}-J_{\mu} = 0$. Therefore the equation of motion of the noncommutative theory in terms of the usual variables,
\begin{equation}
\partial_{0}\partial_{0}A_{\mu}-\boldsymbol{\nabla}^{2}A_{\mu}-J_{\mu}+\uO(\theta) = 0,
\end{equation}
implies
\begin{equation}\label{}
 \left[\partial_{0}\partial_{0}A_{n}(x), J_{0}^5(y)\right] = \boldsymbol{\nabla}_{x}^{2}\left[A_{n}(x), J_{0}^5(y)\right] + \left[J_{n}(x), J_{0}^5(y)\right]+\uO(\theta),
\end{equation}
 which can be computed using Eqs.~\eqref{CS102} and \eqref{CA101}. Thus Eqs.~\eqref{CCA1011}--\eqref{CCA108} become
\begin{gather}
\label{CCB101}\begin{split}
\nc{S}_{00}(x,y) &= \frac{\ui}{4\pi^2}\varepsilon_{ijk}F_{jk}(y)\partial_{i}^{x}\delta^{3}(\mathbf{x}-\mathbf{y})\\
&\quad {}-\frac{\ui}{4\pi^2}\theta^{mn}\varepsilon_{ijk}\left[\partial_{n}^{y}\left(A_{m}(y)F_{jk}(y)\partial_{i}^{x}\delta^{3}(\mathbf{x}-\mathbf{y})\right)\right.\\
&\qquad\qquad\qquad\qquad\left.{}+\partial_{n}^{x}\left(A_{m}(x)F_{jk}(y)\partial_{i}^{x}\delta^{3}(\mathbf{x}-\mathbf{y})\right)\right] + \uO(\theta^2),
\end{split}\displaybreak[0]\\
\label{CCB102} \raisetag{14pt}
\begin{split}
\nc{S}_{i0}(x,y) &= -\frac{\ui}{4\pi^2}\varepsilon_{ijk}F_{0j}(x)\partial_{k}^{y}\delta^{3}(\mathbf{x}-\mathbf{y})\\
&\quad -\frac{\ui}{4\pi^2}\theta^{im}\left[\varepsilon_{njk}\left(F_{m0}(x)F_{jk}(y)\partial_{n}^{x}\delta^{3}(\mathbf{x}-\mathbf{y})+F_{mn}(x)F_{0j}(x)\partial_{k}^{y}\delta^{3}(\mathbf{x}-\mathbf{y})\right)\right.\\
&\qquad\qquad\qquad \left.{}-\varepsilon_{mjk}F_{jk}J_{0}\delta^{3}(\mathbf{x}-\mathbf{y})\right]\\
&\quad +\frac{\ui}{4\pi^2}\theta^{mn}\varepsilon_{ijk}\left[\partial_{n}^{y}\left(A_{m}(y)F_{0j}(x)\partial_{k}^{y}\delta^{3}(\mathbf{x}-\mathbf{y})\right)\right.\\
&\qquad\qquad\qquad\qquad \left.{}+\partial_{n}^{x}\left(A_{m}(x)F_{0j}(x)\partial_{k}^{y}\delta^{3}(\mathbf{x}-\mathbf{y})\right)\right] + \uO(\theta^2),
\end{split}\displaybreak[0]\\
\label{CCB103}\begin{split}
\nc{S}_{0i}(x,y) &= \frac{\ui}{4\pi^2}\varepsilon_{ijk}F_{0j}(y)\partial_{k}^{x}\delta^{3}(\mathbf{x}-\mathbf{y})\\
&\quad - \frac{\ui}{4\pi^2}\theta^{im}\varepsilon_{njk}\left(F_{m0}(y)F_{jk}(y)\partial_{n}^{x}\delta^{3}(\mathbf{x}-\mathbf{y})-F_{mn}(y)F_{0j}(y)\partial_{k}^{x}\delta^{3}(\mathbf{x}-\mathbf{y})\right)\\
&\quad - \frac{\ui}{4\pi^2}\theta^{mn}\varepsilon_{ijk}\left[\partial_{n}^{y}\left(A_{m}(y)F_{0j}(y)\partial_{k}^{x}\delta^{3}(\mathbf{x}-\mathbf{y})\right)\right.\\
&\qquad\qquad\qquad\qquad \left.{} + \partial_{n}^{x}\left(A_{m}(x)F_{0j}(y)\partial_{k}^{x}\delta^{3}(\mathbf{x}-\mathbf{y})\right)\right] + \uO(\theta^2),
\end{split}\displaybreak[0]\\
\label{CCB104}
\nc{L}_{00}(x,y)
= \frac{\ui}{8\pi^2}\theta^{mn}\varepsilon_{njk}A_{m}F_{jk}\delta^{3}(\mathbf{x}-\mathbf{y}) + \uO(\theta^2),\displaybreak[0]\\
\label{CCB105}
\nc{L}_{0i}(x,y)
= \frac{\ui}{8\pi^2}\theta^{mn}\varepsilon_{nik}A_{m}F_{0k}\delta^{3}(\mathbf{x}-\mathbf{y}) + \uO(\theta^2),\displaybreak[0]\\
\label{CCB1051}
\nc{L}_{i0}(x,y)
= \uO(\theta^2),\displaybreak[0]\\
\label{CCB1052}
\nc{L}_{im}(x,y)
= \ui\theta^{mi}J_{0}^{5}\delta^{3}(\mathbf{x}-\mathbf{y}) + \uO(\theta^2),\displaybreak[0]\\
\label{CCB106}\begin{split}
\nc{M}_{00}(x,y)
&= \frac{\ui}{4\pi^2}\theta^{mn}\varepsilon_{njk}\left(\partial_{m}A_{0}F_{jk}\delta^{3}(\mathbf{x}-\mathbf{y}) - \frac{1}{2}A_{m}(x)F_{0j}(x)\partial_{k}^{y}\delta^{3}(\mathbf{x}-\mathbf{y})\right)\\
&\quad {} + \uO(\theta^2),
\end{split}\displaybreak[0]\\
\label{CCB107}\begin{split}
\nc{M}_{i0}(x,y)
&= \frac{\ui}{4\pi^2}\varepsilon_{ijk}F_{jk}\delta^{3}(\mathbf{x}-\mathbf{y}) + \ui\theta^{in}\partial_{n}^{y}\left(J_{0}^{5}\delta^{3}(\mathbf{x}-\mathbf{y})\right)\\
&\quad {} - \frac{\ui}{4\pi^2}\theta^{mn}\bigg[\varepsilon_{mjk}\left\{F_{ni}F_{jk}\delta^{3}(\mathbf{x}-\mathbf{y}) + \frac{1}{2}\partial_{i}^{x}\left(A_{n}F_{jk}\delta^{3}(\mathbf{x}-\mathbf{y})\right)\right\}\\
&\qquad\qquad\qquad {} + \varepsilon_{ijk}A_{m}\partial_{n}F_{jk}\delta^{3}(\mathbf{x}-\mathbf{y})\bigg] + \uO(\theta^2),
\end{split}\displaybreak[0]\\
\label{CCB108}
\begin{split}
\nc{M}_{ik}(x,y)
&= \frac{\ui}{4\pi^2}\varepsilon_{ikj}F_{0j}\delta^{3}(\mathbf{x}-\mathbf{y}) + \ui\theta^{in}\partial_{n}^{y}\left(J_{k}^{5}\delta^{3}(\mathbf{x}-\mathbf{y})\right)+\ui\theta^{ki}J_{n}^{5}(y)\partial_{n}^{y}\delta^{3}(\mathbf{x}-\mathbf{y})\\
&\quad {} -\ui\theta^{km}\left\{J_{i}^{5}(y)\partial_{m}^{y}\delta^{3}(\mathbf{x}-\mathbf{y})+\frac{\ui}{4\pi^2}\varepsilon_{ijn}\left(F_{m0}F_{jn}+F_{mn}F_{0j}\right)\delta^{3}(\mathbf{x}-\mathbf{y})\right\}\\
&\quad {} - \frac{\ui}{4\pi^2}\theta^{mn}\bigg[\varepsilon_{mkj}\left\{F_{ni}F_{0j}\delta^{3}(\mathbf{x}-\mathbf{y}) + \frac{1}{2}\partial_{i}^{x}\left(A_{n}F_{0j}\delta^{3}(\mathbf{x}-\mathbf{y})\right)\right\}\\
&\qquad\qquad\qquad {} + \varepsilon_{ikj}A_{m}\partial_{n}F_{0j}\delta^{3}(\mathbf{x}-\mathbf{y})\bigg] + \uO(\theta^2).
\end{split}
\end{gather}
We have thus obtained various anomalous commutators up to the first order in a magnetic-type~$\theta$. These expressions are given in commutative variables. Using the inverse maps,
\begin{gather}
\label{inv1}
A_{\mu} = \nc{A}_{\mu} + \frac{1}{2}\theta^{\alpha\beta}\nc{A}_{\alpha}\left(\partial_{\beta}\nc{A}_{\mu}+\nc{F}_{\beta\mu}\right) + \uO(\theta^2),\displaybreak[0]\\
\label{inv2}
F_{\mu\nu} = \nc{F}_{\mu\nu} + \theta^{\alpha\beta}\left(\nc{A}_{\alpha}\partial_{\beta}\nc{F}_{\mu\nu} + \nc{F}_{\mu\alpha}\nc{F}_{\beta\nu}\right) + \uO(\theta^2),\displaybreak[0]\\
\label{inv3}
J^{\mu} = \nc{J}^{\mu} + \theta^{\alpha\beta}\left(\nc{A}_{\alpha}\partial_{\beta}\nc{J}^{\mu} - \frac{1}{2}\nc{F}_{\alpha\beta}\nc{J}^{\mu}\right)-\theta^{\mu\alpha}\nc{F}_{\alpha\beta}\nc{J}^{\beta} + \uO(\theta^2),
\end{gather}
with $\theta^{0i} = 0$, we can express them in terms of the noncommutative variables:
\begin{gather}
\label{CCC101} \raisetag{18pt}
\begin{split}
\nc{S}_{00}(x,y)
&= \frac{\ui}{4\pi^2}\varepsilon_{ijk}\nc{F}_{jk}(y)\partial_{i}^{x}\delta^{3}(\mathbf{x}-\mathbf{y})\\
&\quad {}+\frac{\ui}{4\pi^2}\theta^{mn}\varepsilon_{ijk}\left[\nc{F}_{jm}(y)\nc{F}_{nk}(y)\partial_{i}^{x}\delta^{3}(\mathbf{x}-\mathbf{y})-\nc{F}_{jk}(y)\partial_{n}^{y}\left(\nc{A}_{m}(y)\partial_{i}^{x}\delta^{3}(\mathbf{x}-\mathbf{y})\right)\right.\\
&\qquad\qquad\qquad\qquad \left.{}-\nc{F}_{jk}(y)\partial_{n}^{x}\left(\nc{A}_{m}(x)\partial_{i}^{x}\delta^{3}(\mathbf{x}-\mathbf{y})\right)\right] + \uO(\theta^2),
\end{split}\displaybreak[0]\\
\label{CCC102} \raisetag{18pt}
\begin{split}
\nc{S}_{i0}(x,y)
&= -\frac{\ui}{4\pi^2}\varepsilon_{ijk}\nc{F}_{0j}(x)\partial_{k}^{y}\delta^{3}(\mathbf{x}-\mathbf{y})\\
&\quad {} - \frac{\ui}{4\pi^2}\theta^{im}\left[\varepsilon_{njk}\left(\nc{F}_{m0}(x)\nc{F}_{jk}(y)\partial_{n}^{x}\delta^{3}(\mathbf{x}-\mathbf{y})+\nc{F}_{mn}(x)\nc{F}_{0j}(x)\partial_{k}^{y}\delta^{3}(\mathbf{x}-\mathbf{y})\right)\right.\\
&\qquad\qquad\qquad \left.{} - \varepsilon_{mjk}\nc{J}_{0}\nc{F}_{jk}\delta^{3}(\mathbf{x}-\mathbf{y})\right]\\
&\quad {} - \frac{\ui}{4\pi^2}\theta^{mn}\varepsilon_{ijk}\left[\nc{F}_{0m}(x)\nc{F}_{nj}(x)\partial_{k}^{y}\delta^{3}(\mathbf{x}-\mathbf{y})-\nc{F}_{0j}(x)\partial_{n}^{y}\left(\nc{A}_{m}(y)\partial_{k}^{y}\delta^{3}(\mathbf{x}-\mathbf{y})\right)\right.\\
&\qquad\qquad\qquad\qquad \left.{} - \nc{F}_{0j}(x)\partial_{n}^{x}\left(\nc{A}_{m}(x)\partial_{k}^{y}\delta^{3}(\mathbf{x}-\mathbf{y})\right)\right] + \uO(\theta^2),
\end{split}\displaybreak[0]\\
\label{CCC103} \raisetag{18pt}
\begin{split}
\nc{S}_{0i}(x,y)
&= \frac{\ui}{4\pi^2}\varepsilon_{ijk}\nc{F}_{0j}(y)\partial_{k}^{x}\delta^{3}(\mathbf{x}-\mathbf{y})\\
&\quad {} - \frac{\ui}{4\pi^2}\theta^{im}\varepsilon_{njk}\left(\nc{F}_{m0}(y)\nc{F}_{jk}(y)\partial_{n}^{x}\delta^{3}(\mathbf{x}-\mathbf{y}) - \nc{F}_{mn}(y)\nc{F}_{0j}(y)\partial_{k}^{x}\delta^{3}(\mathbf{x}-\mathbf{y})\right)\\
&\quad {} + \frac{\ui}{4\pi^2}\theta^{mn}\varepsilon_{ijk}\left[\nc{F}_{0m}(y)\nc{F}_{nj}(y)\partial_{k}^{x}\delta^{3}(\mathbf{x}-\mathbf{y}) - \nc{F}_{0j}(y)\partial_{n}^{y}\left(\nc{A}_{m}(y)\partial_{k}^{x}\delta^{3}(\mathbf{x}-\mathbf{y})\right)\right.\\
&\qquad\qquad\qquad\qquad \left.{} - \nc{F}_{0j}(y)\partial_{n}^{x}\left(\nc{A}_{m}(x)\partial_{k}^{x}\delta^{3}(\mathbf{x}-\mathbf{y})\right)\right] + \uO(\theta^2),
\end{split}\displaybreak[0]\\
\label{CCC104}
\nc{L}_{00}(x,y)
= \frac{\ui}{8\pi^2}\theta^{mn}\varepsilon_{njk}\nc{A}_{m}\nc{F}_{jk}\delta^{3}(\mathbf{x}-\mathbf{y}) + \uO(\theta^2),\displaybreak[0]\\
\label{CCC105}
\nc{L}_{0i}(x,y)
= \frac{\ui}{8\pi^2}\theta^{mn}\varepsilon_{nik}\nc{A}_{m}\nc{F}_{0k}\delta^{3}(\mathbf{x}-\mathbf{y}) + \uO(\theta^2),\displaybreak[0]\displaybreak[0]\\
\label{CCC1051}
\nc{L}_{i0}(x,y)
= \uO(\theta^2),\displaybreak[0]\\
\label{CCC1052}
\nc{L}_{im}(x,y)
= \ui\theta^{mi}\nc{J}_{0}^{5}\delta^{3}(\mathbf{x}-\mathbf{y}) + \uO(\theta^2),\displaybreak[0]\\
\label{CCC106}\begin{split}
\nc{M}_{00}(x,y)
&= \frac{\ui}{4\pi^2}\theta^{mn}\varepsilon_{njk}\left(\partial_{m}\nc{A}_{0}\nc{F}_{jk}\delta^{3}(\mathbf{x}-\mathbf{y}) - \frac{1}{2}\nc{A}_{m}(x)\nc{F}_{0j}(x)\partial_{k}^{y}\delta^{3}(\mathbf{x}-\mathbf{y})\right)\\
&\quad {} + \uO(\theta^2),
\end{split}\displaybreak[0]\\
\label{CCC107}\begin{split}
\nc{M}_{i0}(x,y)
&= \frac{\ui}{4\pi^2}\varepsilon_{ijk}\nc{F}_{jk}\delta^{3}(\mathbf{x}-\mathbf{y}) + \ui\theta^{in}\partial_{n}^{y}\left(\nc{J}_{0}^{5}\delta^{3}(\mathbf{x}-\mathbf{y})\right)\\
&\quad {} - \frac{\ui}{4\pi^2}\theta^{mn}\bigg[\varepsilon_{mjk}\left\{\nc{F}_{ni}\nc{F}_{jk}\delta^{3}(\mathbf{x}-\mathbf{y}) + \frac{1}{2}\partial_{i}^{x}\left(\nc{A}_{n}\nc{F}_{jk}\delta^{3}(\mathbf{x}-\mathbf{y})\right)\right\}\\
&\qquad\qquad\qquad {} - \varepsilon_{ijk}\nc{F}_{jm}\nc{F}_{nk}\delta^{3}(\mathbf{x}-\mathbf{y})\bigg] + \uO(\theta^2),
\end{split}\displaybreak[0]\\
\label{CCC108} \raisetag{21pt}
\begin{split}
\nc{M}_{ik}(x,y)
&= \frac{\ui}{4\pi^2}\varepsilon_{ikj}\nc{F}_{0j}\delta^{3}(\mathbf{x}-\mathbf{y}) + \ui\theta^{in}\partial_{n}^{y}\left(\nc{J}_{k}^{5}\delta^{3}(\mathbf{x}-\mathbf{y})\right) + \ui\theta^{ki}\nc{J}_{n}^{5}(y)\partial_{n}^{y}\delta^{3}(\mathbf{x}-\mathbf{y})\\
&\quad {} - \ui\theta^{km}\left\{\nc{J}_{i}^{5}(y)\partial_{m}^{y}\delta^{3}(\mathbf{x}-\mathbf{y})+\frac{\ui}{4\pi^2}\varepsilon_{ijn}\left(\nc{F}_{m0}\nc{F}_{jn}+\nc{F}_{mn}\nc{F}_{0j}\right)\delta^{3}(\mathbf{x}-\mathbf{y})\right\}\\
&\quad {} - \frac{\ui}{4\pi^2}\theta^{mn}\bigg[ \varepsilon_{mkj}\left\{\nc{F}_{ni}\nc{F}_{0j}\delta^{3}(\mathbf{x}-\mathbf{y}) + \frac{1}{2}\partial_{i}^{x}\left(\nc{A}_{n}\nc{F}_{0j}\delta^{3}(\mathbf{x}-\mathbf{y})\right)\right\}\\
&\qquad\qquad\qquad {} - \varepsilon_{ikj}\nc{F}_{0m}\nc{F}_{nj}\delta^{3}(\mathbf{x}-\mathbf{y})\bigg] + \uO(\theta^2).
\end{split}
\end{gather}
This completes our obtention of the anomalous commutators in both commutative as well as noncommutative variables.


\section{\label{sec:cons}Consistency conditions and the anomalous commutators}

Just as the anomalous commutators in the usual theory are subjected to certain consistency conditions \cite{AB}, we now show that those in the noncommutative theory also obey certain consistency conditions, implying their compatibility with the noncommutative covariant anomaly~\eqref{anom-cov.d}.

To obtain the consistency criteria, we begin with
\begin{equation}\label{cons101}
\partial_{0}\nc{S}_{00}(x,y) = \partial_{0}\left[\nc{J}_{0}(x), \nc{J}_{0}^5(y)\right]= \left[\partial_{0}\nc{J}_{0}(x), \nc{J}_{0}^5(y)\right]+\left[\nc{J}_{0}(x), \partial_{0}\nc{J}_{0}^5(y)\right].
\end{equation}
In view of Eq.~\eqref{DJ}, it follows from $\nc{\uD}_{\mu}\star\nc{J}^{\mu} = 0$, and  $\nc{\uD}_{\mu}\star\nc{J}^{\mu}_{5} = \nc{\mathscr{A}}$ that (for $\theta^{0i}=0$)
\begin{gather}
\label{cons102}
\partial_{0}\nc{J}_{0} = \partial_{m}\nc{J}_{m}+\theta^{mn}\partial_{m}\nc{J}^{\mu}\partial_{n}\nc{A}_{\mu}+\uO(\theta^2),\\
\label{cons103}
\partial_{0}\nc{J}_{0}^5 = \partial_{m}\nc{J}_{m}^{5}+\theta^{mn}\partial_{m}\nc{J}^{\mu}_{5}\partial_{n}\nc{A}_{\mu}+\nc{\mathscr{A}}+\uO(\theta^2).
\end{gather}
Using these to substitute for $\partial_{0}\nc{J}_{0}$ and $\partial_{0}\nc{J}_{0}^5$, Eq.~\eqref{cons101} yields a consistency relation among the anomalous commutators of the noncommutative theory:
\begin{equation}\label{cons104}\begin{split}
\partial_{0}\nc{S}_{00}(x,y) &= \partial_{m}^{x}\nc{S}_{m0}(x,y)+\partial_{m}^{y}\nc{S}_{0m}(x,y)\\
&\quad +\theta^{mn}\left(\partial_{n}\nc{A}^{\mu}(x)\partial_{m}^{x}\nc{S}_{\mu0}(x,y)+\partial_{n}\nc{A}^{\mu}(y)\partial_{m}^{y}\nc{S}_{0\mu}(x,y)\right.\\
&\qquad\qquad \left.{} + \partial_{m}\nc{J}^{\mu}(x)\partial_{n}^{x}\nc{L}_{\mu 0}(x,y) + \partial_{m}\nc{J}^{\mu}_{5}(y)\partial_{n}^{y}\left[\nc{J}_{0}(x), \nc{A}_{\mu}(y)\right]\right)\\
&\quad {} + \left[\nc{J}_{0}(x), \nc{\mathscr{A}}(y)\right]+\uO(\theta^2).
\end{split}\end{equation}
The essentially new ingredient is the last bracket involving the anomaly. Using the maps (with $\theta^{0i} = 0$) for $\nc{J}_{0}$ and $\nc{\mathscr{A}}$ given in Eqs.~\eqref{J-hat2.m} and \eqref{anom.m} respectively, we get
\begin{equation*}\begin{split}
\left[\nc{J}_{0}(x), \nc{\mathscr{A}}(y)\right] &= \left[J_{0}(x), \mathscr{A}(y)\right]\\
&\quad {} - \theta^{mn}\left(\partial_{n}^{y}\left[J_{0}(x), A_{m}(y)\mathscr{A}(y)\right] + \partial_{n}^{x}\left[A_{m}(x)J_{0}(x), \mathscr{A}(y)\right]\right)+\uO(\theta^2),
\end{split}
\end{equation*}
which, on substituting for the anomaly, $\mathscr{A} = ({1}/{16\pi^{2}})\varepsilon_{\mu\nu\lambda\rho}F^{\mu\nu}F^{\lambda\rho}$, 
and using the relations \eqref{Scr} and \eqref{SCC}, yields
\begin{equation}\label{cons105}\begin{split}
\left[\nc{J}_{0}(x), \nc{\mathscr{A}}(y)\right] = \frac{\ui}{4\pi^2}\theta^{mn}\varepsilon_{mjk}J_{0}(y)F_{jk}(y)\partial_{n}^{x}\delta^{3}(\mathbf{x}-\mathbf{y})+\uO(\theta^2).
\end{split}
\end{equation}
We observe that the $\theta\rightarrow 0$ limit of the condition \eqref{cons104} is
\begin{equation}\label{cons1041}
\partial_{0}S_{00}(x,y) = \partial_{m}^{x}S_{m0}(x,y) + \partial_{m}^{y}S_{0m}(x,y),
\end{equation}
which is easily verified using Eqs.~\eqref{CS101}--\eqref{CS103}. To show that Eq.~\eqref{cons104} indeed holds is also straight-forward. Equation~\eqref{cons105} gives the last term on the right-hand side of Eq.~\eqref{cons104}. The commutator $[\nc{J}_{0}(x), \nc{A}_{\mu}(y)]$ occurs in an $\uO(\theta)$ term, and therefore it can be replaced by $[J_{0}(x), A_{\mu}(y)]$ which vanishes because of Eq.~\eqref{SCC}. The other terms in Eq.~\eqref{cons104} are also known in view of Eqs.~\eqref{CCB101}--\eqref{CCB108}. Substituting for all these commutators, we find that Eq.~\eqref{cons104} is satisfied. Alternatively, the verification of Eq.~\eqref{cons104} can be done in noncommutative variables by exploiting Eqs.~\eqref{CCC101}--\eqref{CCC108} and the one obtained by using the inverse maps \eqref{inv2} and \eqref{inv3} on the right-hand side of Eq.~\eqref{cons105} (this amounts to just replacing the usual variables by the noncommutative ones, since it is already an $\uO(\theta)$ term). This shows that our anomalous commutators are compatible with the noncommutative anomaly.

As another example of a consistency condition, we note that
\begin{equation*}
\partial_{0}\left[\nc{A}_{\nu}(x), \nc{J}_{0}^5(y)\right] = \left[\partial_{0}\nc{A}_{\nu}(x), \nc{J}_{0}^5(y)\right]+\left[\nc{A}_{\nu}(x), \partial_{0}\nc{J}_{0}^5(y)\right],
\end{equation*}
which, invoking the notations introduced earlier, can be rewritten compactly as
\begin{equation}\label{consc}
\partial_{0}\nc{L}_{\nu 0}(x,y) = \nc{M}_{\nu 0}(x,y) + \left[\nc{A}_{\nu}(x), \partial_{0}\nc{J}_{0}^5(y)\right].
\end{equation}
Using Eq.~\eqref{cons103} to substitute for $\partial_{0}\nc{J}_{0}^5$ on the right-hand side gives a consistency condition
\begin{equation}\label{cons107}\begin{split}
\partial_{0}\nc{L}_{\nu 0}(x,y) &= \nc{M}_{\nu 0}(x,y) + \partial_{m}^{y}\nc{L}_{\nu m}(x,y)\\
&\quad {}+\theta^{mn}\left(\partial_{n}\nc{A}^{\mu}(y)\partial_{m}^{y}\nc{L}_{\nu\mu}(x,y)
+\partial_{m}\nc{J}^{\mu}_{5}(y)\partial_{n}^{y}\left[\nc{A}_{\nu}(x), \nc{A}_{\mu}(y)\right]\right)\\
&\quad {}+\left[\nc{A}_{\nu}(x), \nc{\mathscr{A}}(y)\right] + \uO(\theta^2).
\end{split}\end{equation}
Using the maps for $\nc{A}_{0}$ and $\nc{\mathscr{A}}$ given in Eqs.~\eqref{A.m} and \eqref{anom.m} we get
\begin{equation*}\begin{split}
\left[\nc{A}_{0}(x), \nc{\mathscr{A}}(y)\right] &= \left[A_{0}(x), \mathscr{A}(y)\right] - \theta^{mn}\bigg(\frac{1}{2}\left[A_{m}(x)(\partial_{n}A_{0}(x)+F_{n0}(x)), \mathscr{A}(y)\right]\\
&\qquad\qquad\qquad\qquad\qquad\quad {} + \partial_{n}^{y}\left[A_{0}(x), A_{m}(y)\mathscr{A}(y)\right]\bigg) + \uO(\theta^2).
\end{split}\end{equation*}
By substituting for the anomaly, $\mathscr{A}= ({1}/{16\pi^{2}})\varepsilon_{\mu\nu\lambda\rho}F^{\mu\nu}F^{\lambda\rho}$, and using Eq.~\eqref{Scr}, this is computed as
\begin{equation}\label{cons113}\begin{split}
\left[\nc{A}_{0}(x), \nc{\mathscr{A}}(y)\right] &= \frac{\ui}{4\pi^2}\theta^{mn}\varepsilon_{mjk}\bigg[\frac{1}{2}\left(\partial_{n}A_{0}+F_{n0}\right)F_{jk}\delta^{3}(\mathbf{x}-\mathbf{y})\\
&\qquad\qquad\qquad\quad {} - F_{0j}(y)A_{n}(x)\partial_{k}^{y}\delta^{3}(\mathbf{x}-\mathbf{y})\bigg] + \uO(\theta^2).
\end{split}\end{equation}
Similarly we get
\begin{equation}
\label{cons108}\begin{split}
\left[\nc{A}_{i}(x), \nc{\mathscr{A}}(y)\right] &= -\frac{\ui}{4\pi^2}\varepsilon_{ijk}F_{jk}\delta^{3}(\mathbf{x}-\mathbf{y})\\
&\quad {}+\frac{\ui}{4\pi^2}\theta^{mn}\bigg[\varepsilon_{mjk}\left\{F_{ni}F_{jk}\delta^{3}(\mathbf{x}-\mathbf{y}) + \frac{1}{2}\partial_{i}^{x}\left(A_{n}F_{jk}\delta^{3}(\mathbf{x}-\mathbf{y})\right)\right\}\\
&\qquad\qquad\qquad {} + \varepsilon_{ijk}\left(A_{m}\partial_{n}F_{jk}\delta^{3}(\mathbf{x}-\mathbf{y})\right)\bigg] + \uO(\theta^2).
\end{split}
\end{equation}
Also, in view of the map \eqref{A.m}, we observe that $[\nc{A}_{\nu}(x), \nc{A}_{\mu}(y)]$ will not have at least any $\theta$-independent part, which means that the term involving this commutator on the right-hand side of Eq.~\eqref{cons107} drops out. Using Eqs.~\eqref{A.m}, \eqref{CCB104}, \eqref{CCB105}, \eqref{CCB106} and \eqref{cons113}, the right-hand side of Eq.~\eqref{cons107} for $\nu = 0$ reduces to
\begin{equation}\label{cons114}
\frac{\ui}{8\pi^2}\theta^{mn}\varepsilon_{njk}\partial_{0}\left(A_{m}F_{jk}\right)\delta^{3}(\mathbf{x}-\mathbf{y}) + \uO(\theta^2),
\end{equation}
which is also what the left-hand side of Eq.~\eqref{cons107} for $\nu = 0$ reduces to upon substituting for the commutator from Eq.~\eqref{CCB104}. For $\mu = i$, the left-hand side of Eq.~\eqref{cons107}, up to $\uO(\theta)$, vanishes in view of the Eq.~\eqref{CCB1051}, and the right-hand side, using Eqs.~\eqref{A.m}, \eqref{CCB1051}, \eqref{CCB1052}, \eqref{CCB107} and \eqref{cons108}, also vanishes. This shows the compatibility of the noncommutative anomalous commutators with the noncommutative anomaly.


\section{\label{sec:ambig}Ambiguities in anomalous commutators and the consistency conditions}

As mentioned in Ref.~\cite{AB}, the commutators given in the set \eqref{CS101}--\eqref{CB103} for the ordinary theory have been deduced from the triangle graph alone, which is also responsible for the current-divergence anomaly. This does not rule out the possibility that higher orders of perturbation theory may modify the values of these commutators. However, the commutators $S_{00}(x,y)$ and $M_{i0}(x,y)$ can also be deduced from simpler, exact commutators and equations of motion, which suggests that their value is exact to all orders of perturbation theory. On the other hand, the values given in the set \eqref{CS101}--\eqref{CB103} for the commutators $S_{i0}(x,y)$, $S_{0i}(x,y)$ and $M_{ik}(x,y)$ cannot be deduced in a way similar to those of $S_{00}(x,y)$ and $M_{i0}(x,y)$, and the possibility of the presence of additional terms is not ruled out. It has been shown \cite{AB} that if values of these commutators are modified to
\begin{gather}
\label{CS102m}
S_{i0}(x,y)
= -\frac{\ui}{4\pi^2}\varepsilon_{ijk}F_{0j}(x)\partial_{k}^y\delta^{3}(\mathbf{x}-\mathbf{y})+\ui\partial_{k}^{y}\left(T^{ik}\delta^{3}(\mathbf{x}-\mathbf{y})\right),\displaybreak[0]\\
\label{CS103m}
S_{0i}(x,y)
= \frac{\ui}{4\pi^2}\varepsilon_{ijk}F_{0j}(y)\partial_{k}^x\delta^{3}(\mathbf{x}-\mathbf{y})-\ui\partial_{k}^{x}\left(T^{ki}\delta^{3}(\mathbf{x}-\mathbf{y})\right),\displaybreak[0]\\
\label{CB103m}
M_{im}(x,y)
= \frac{\ui}{4\pi^2}\varepsilon_{imn}F_{0n}\delta^{3}(\mathbf{x}-\mathbf{y})-\ui T^{im}\delta^{3}(\mathbf{x}-\mathbf{y}),
\end{gather}
with $T^{ik}(y)$ a pseudotensor operator, then the consistency conditions, Eq.~\eqref{cons1041} for example, are unchanged. The implications of these modifications will now be analysed in the present context.

The first point to note is that the various anomalous commutators might get altered due to the additional $T^{ij}$-dependent pieces. We explicitly compute these modifications. Equations~\eqref{CCA1011}--\eqref{CCA108} relate the anomalous commutators in the noncommutative theory with their commutative counterparts. It becomes clear from these equations that the modifications \eqref{CS102m}--\eqref{CB103m} will not alter  the values of the commutators $\nc{S}_{00}(x,y)$, $\nc{L}_{00}(x,y)$, $\nc{L}_{i0}(x,y)$, $\nc{L}_{im}(x,y)$ and $\nc{M}_{i0}(x,y)$ as given in the set \eqref{CCB101}--\eqref{CCB108}. The values of the remaining commutators will be modified as
\begin{gather}
\label{amb101}\begin{split}
\nc{S}_{i0}(x,y) &= (\text{right-hand side of Eq.~\eqref{CCB102}}) + \ui\partial_{k}^{y}\left(T^{ik}\delta^{3}(\mathbf{x}-\mathbf{y})\right)\\
&\quad {} + \ui\theta^{im}F_{mn}(x)\partial_{k}^{y}\left(T^{nk}\delta^{3}(\mathbf{x}-\mathbf{y})\right)\\
&\quad {} - \ui\theta^{mn}\left[\partial_{n}^{y}\left\{A_{m}(y)\partial_{k}^{y}\left(T^{ik}\delta^{3}(\mathbf{x}-\mathbf{y})\right)\right\}\right.\\
&\qquad\qquad\quad \left.{} + \partial_{n}^{x}\left\{A_{m}(x)\partial_{k}^{y}\left(T^{ik}\delta^{3}(\mathbf{x}-\mathbf{y})\right)\right\}\right],
\end{split}\displaybreak[0]\\
\label{amb102}\begin{split}
\nc{S}_{0i}(x,y) &= (\text{right-hand side of Eq.~\eqref{CCB103}}) - \ui\partial_{k}^{x}\left(T^{ki}\delta^{3}(\mathbf{x}-\mathbf{y})\right)\\
&\quad {} - \ui\theta^{im}F_{mn}(y)\partial_{k}^{x}\left(T^{kn}\delta^{3}(\mathbf{x}-\mathbf{y})\right)\\
&\quad {}+\ui\theta^{mn}\left[\partial_{n}^{y}\left\{A_{m}(y)\partial_{k}^{x}\left(T^{ki}\delta^{3}(\mathbf{x}-\mathbf{y})\right)\right\}\right.\\
&\qquad\qquad\quad \left.{} + \partial_{n}^{x}\left\{A_{m}(x)\partial_{k}^{x}\left(T^{ki}\delta^{3}(\mathbf{x}-\mathbf{y})\right)\right\}\right],
\end{split}\displaybreak[0]\\
\label{amb103}
\nc{L}_{0i}(x,y) = (\text{right-hand side of Eq.~\eqref{CCB105}}) -\frac{\ui}{2}\theta^{mn}A_{m}T^{ni}\delta^{3}(\mathbf{x}-\mathbf{y}),\displaybreak[0]\\
\label{amb104}
\nc{M}_{00}(x,y) = (\text{right-hand side of Eq.~\eqref{CCB106}}) +\frac{\ui}{2}\theta^{mn}A_{m}(x)\partial_{i}^{y}\left(T^{ni}\delta^{3}(\mathbf{x}-\mathbf{y})\right),\displaybreak[0]\\
\label{amb105}
\nc{M}_{ik}(x,y) = (\text{right-hand side of Eq.~\eqref{CCB108}}) + (\cdots),
\end{gather}
where $(\cdots)$ appearing on the right-hand side of Eq.~\eqref{amb105} represents the terms involving $T^{ik}(y)$ whose explicit structure is not needed for our purpose.

Next we show that the conditions \eqref{cons104} and \eqref{cons107} still hold. The left-hand side of the condition~\eqref{cons104} does not involve any of the modified commutators  given in the set \eqref{amb101}--\eqref{amb105}, its value therefore remains unaltered. The right-hand side does involve the modified commutators, but it is a matter of straight-forward algebra to show that there is no change in its value. The consistency condition \eqref{cons107} for $\nu = i$ does not involve any of the modified commutators, and therefore it trivially remains valid. As far as the condition \eqref{cons107} with $\nu = 0$ is concerned, its left-hand side is $\partial_{0}\nc{L}_{00}(x,y)$ whose value obviously remains unaffected. The right-hand side involves the modified commutators, but again after some algebra we find that its value remains unchanged.


\section{\label{sec:conclu}Conclusions}

We have obtained the $\uO(\theta)$ structure of all the anomalous commutators involving the covariant axial-vector current in noncommutative electrodynamics for a magnetic-type $\theta$. The basic step in our approach is to exploit the SW maps for currents and fields that relate the noncommutative and usual (commutative) descriptions. The commutators in the noncommutative theory are thereby expressed in terms of their commutative counterparts which are known. Substituting for these known commutators we obtained the commutators in the noncommutative theory. The results were displayed both in terms of the commutative (usual) and noncommutative variables.

One might be tempted to guess the structures of these anomalous commutators as those obtained by a naive covariant deformation of the ordinary results, just as the covariant divergence anomaly~\eqref{anom-cov.d} is obtained by a covariant deformation of the usual result~\eqref{anom.d}. But a simple inspection rules out this possibility. The point is that the covariant deformation of a gauge-invariant expression can only give a star-gauge-covariant expression. Since the currents $J^{\mu}_{5}$ and $\nc{J}^{\mu}_{5}$ are, respectively, gauge invariant and star-gauge covariant, so are the divergences $\partial_{\mu}J^{\mu}_{5}$ and $\nc{\uD}_{\mu}\star\nc{J}^{\mu}_{5}$. One could therefore expect that the star-gauge-covariant anomaly is obtained by a covariant deformation of the usual gauge-invariant anomaly. Explicit calculations serve to verify this expectation~\cite{AS-GM}. On the other hand, although the commutator $[J_{0}(x), J_{0}^{5}(y)]$, for example, is gauge invariant, yet its noncommutative counterpart, $[\nc{J}_{0}(x), \nc{J}_{0}^{5}(y)]$, is not star-gauge covariant because it involves two distinct spacetime points, $x$ and $y$. Therefore it becomes clear that the non-covariant commutator, $[\nc{J}_{0}(x), \nc{J}_{0}^{5}(y)]$, cannot be obtained by just a standard covariant deformation of the usual gauge-invariant commutator. Equations \eqref{CCC101}--\eqref{CCC108} indeed show that there is a departure from the naive covariant deformation of the corresponding gauge-invariant expressions.

We have shown that the commutators we obtained are compatible with the noncommutative covariant anomaly. For this we derived certain consistency conditions involving this anomaly and then showed that the commutators indeed satisfy these conditions. It may be remarked that such consistency conditions were used in usual electrodynamics to reveal the compatibility of the various anomalous commutators with the ABJ anomaly. In the usual QED without axial-vector currents, anomalies in potential--current commutators (`seagulls') and in current--current commutators (`Schwinger terms') are related and cancel exactly when the divergence of covariant matrix element is taken, reproducing the familiar current conservation. The distinguishing feature of the commutator anomalies associated with the triangle diagram is that when the axial-vector divergence is taken, the seagulls and Schwinger terms do not cancel~\cite{Jackiw:1968}. Rather, they combine to give the divergence anomaly (ABJ anomaly), giving an alternative interpretation of the divergence anomaly as the result of non-cancellation of seagulls and Schwinger terms. Our analysis thus suggests that the star-gauge-covariant anomaly can also be regarded as consequence of a similar effect in noncommutative electrodynamics.  Finally, we analysed the implications of certain ambiguities present in the ordinary commutators on our scheme, and showed that the commutators satisfy the consistency conditions irrespective of these ambiguities.

The implications of SW maps were discussed previously~\cite{BLY, BK} in the context of divergence anomalies. Here we find that these maps are also useful in obtaining commutator anomalies. Although we analysed the case of the star-gauge-covariant current, it should be possible to extend this analysis to the star-gauge-invariant current since corresponding SW maps are known to exist~\cite{BG}.


\section*{Acknowledgments}

K.~K.~thanks the Council of Scientific and Industrial Research (CSIR), Government of India for financial support.



\end{document}